\documentstyle[preprint,floats,epsfig,pra,aps]{revtex}
\tightenlines
\begin{document}

\title{Wigner rotations in laser cavities}

\author{S. Ba{\c s}kal \footnote{electronic address:
baskal@newton.physics.metu.edu.tr}}

\address{Department of Physics, Middle East Technical University,
06531 Ankara, Turkey \thanks{Permanent address} \\
Department of Physics, University of Maryland,
College Park, Maryland 20742}

\author {Y. S. Kim \footnote{electronic address: yskim@physics.umd.edu}}
\address{ Department of Physics, University of Maryland, College Park,
Maryland 20742}

\maketitle

\begin{abstract}
The Wigner rotation is a key word in many branches of physics,
chemistry and engineering sciences.  It is a group theoretical effect
resulting from two Lorentz boosts.  The net effect is one boost followed
or preceded by a rotation.  This rotation can therefore be formulated
as a product of three boosts.  In relativistic kinematics, it is a
rotation in the Lorentz frame where the particle is at rest.  This
rotation does not change its momentum, but it rotates the direction
of the spin.  The Wigner rotation is not confined to relativistic
kinematics.  It manifests itself in physical systems where the
underlying mathematics is the Lorentz group.  It is by now widely
known that this group is the basic scientific language for quantum
and classical optics.  It is shown that optical beams perform Wigner
rotations in laser cavities.

\end{abstract}


\narrowtext

The Wigner rotation is a kinematical effect resulting from two
successive Lorentz boosts along different directions.  The result is
not another Lorentz boost, but a boost followed by a rotation.
This rotation is commonly called the Wigner rotation.  In his 1939
paper on the Lorentz group~\cite{wig39}, Wigner indeed emphasized
the importance of the rotation subgroup of the Lorentz group and its
physical significance. Since then, the word ``Wigner rotation''
mentioned frequently in many branches of physics.

The earliest manifestation of the Wigner rotation is the Thomas
precession which we observe in atomic spectra.  Thomas formulated
this problem thirteen years before the appearance of Wigner's 1939
paper~\cite{thomas26}.  The Thomas effect in nuclear spectroscopy
is mentioned in Jackson's book on electrodynamics~\cite{jack99}.
Recently, as the relativistic effects play more prominent roles, the
Wigner rotation has become of the key issues in field theory of
extended objects~\cite{yama81}, electron beams~\cite{cioc92},
relativistic quark model~\cite{ma98,wamb95}, nuclear
scattering~\cite{tjon91}, neutrino physics~\cite{baren96}, as well
as many other areas of physics, chemistry and engineering
sciences~\cite{chen86}.

It is important to note that special relativity is not the only field
of physics where the Lorentz group plays as the fundamental scientific
language.  For instance, in the physics of phase space,
the symmetry group governing linear canonical transformations is the
symplectic group which consists of rotations and squeeze operations.
For the two-dimensional phase space consisting of one coordinate
and one momentum variables, the group governing linear canonical
transformation the symplectic group $Sp(2)$.  This group is locally
isomorphic to the Lorentz group $O(2,1)$ applicable to two space and
one time-dimensions. The squeeze transformation in phase space is like
the Lorentz boost in special relativity.  Here also we can consider
two successive squeezes which result in one squeeze followed by a
rotation.  This is clearly another form of the Wigner
rotation~\cite{hhk89}.  The physics of phase space covers not only
classical mechanics but also squeezed states of light~\cite{knp91}

Another recent trend is that the Lorentz group is becoming prominent
in classical optics, including polariazation optics~\cite{hkn99},
interferometers~\cite{hkn00}, multilayer optics~\cite{monz01,georg01}.
As for lens optics, the formalism starts with two-by-two matrices
representing a lens with its focal length and a translation.  Repeated
applications of these matrices lead to a two-by-two matrix representing
the $Sp(2)$.  Thus, the fundamental scientific language in lens optics
is clearly the group $Sp(2)$~\cite{sudar85,baskal01}.  Thus, it would
not be surprising to see another form of the Wigner rotation in lens
optics.

Let us note that the geomentrical optics of laser cavities is a from
of lens optics.  In this paper, we would like to report that light
waves in a laser cavity are performing Wigner rotations.  We consider
in this paper a cavity bounded by two identical mirrors.  Then the
problem can be translated into an optical system consisting of a
chains of identical lenses separated by the same distance.  One
complete cycle consists of two lenses and two translations.  We shall
show that this complete cycle performs two repeated Wigner rotations.

\begin{figure}[thb]
\begin{center}
\epsfysize=65mm \epsffile{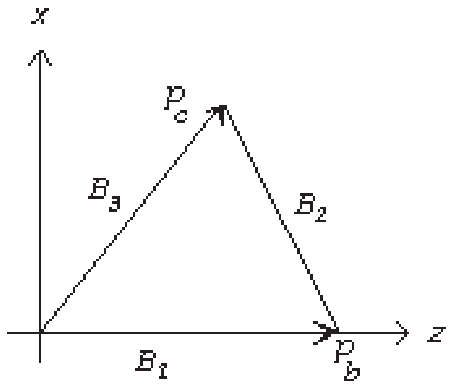}
\vspace{3mm}
\caption{Closed Lorentz boosts. Initially, a massive particle is at
rest with its four momentum $P_{a}$.  The first boost $B_{1}$ brings
$P_{a}$ to $P_{b}$.  The second boost $B_{2}$ transforms $P_{b}$ to
$P_{c}$.  The third boost $B_{3}$ brings $P_{c}$ back to $P_{a}$.
The particle is again at rest. The net effect is a rotation
around the axis perpendicular to the plane containing these three
transformations.  We may assume for convenience that $P_{b}$ is
along the $z$ axis, and $P_{c}$ in the $zx$ plane. The rotation
is then made around the $y$ axis.} \label{wr11}
\end{center}
\end{figure}

For this purpose, let us define precisely the Wigner rotation.  This
rotation is necessary because a product of two boost matrices in
different directions is a boost followed or preceded by a rotation
matrix.  Here, there are three boosts and one rotation.  Thus, the
simplest way to construct a Wigner rotation is to arrange three
boost matrices leading to a rotation matrix~\cite{hks87cqg}.  For
this purpose, let us perform three boosts as illustrated in
Fig.~\ref{wr11}.

Let us start with a particle at rest, with its four momentum
\begin{equation}\label{4a}
P_{a} = (m, 0, 0, 0),
\end{equation}
where we use the metric convention $(ct, z, x, y)$.  Let us next boost this
four-momentum along the $z$ direction using the matrix
\begin{equation}\label{boost1}
B_{1} = \pmatrix{\cosh\eta & \sinh\eta & 0 & 0 \cr
 \sinh\eta & \cosh\eta & 0& 0 \cr 0 & 0 & 1 &  \cr 0 & 0 & 0 & 1} ,
\end{equation}
resulting in the four-momentum
\begin{equation}\label{4b}
P_{b} = m (\cosh\eta, \sinh\eta, 0, 0) .
\end{equation}

Let us rotate this vector around the $y$ axis by an angle $\theta$.
Then the resulting four-momentum is
\begin{equation}\label{4c}
P_{c} = m \left(\cosh\eta, (\sinh\eta)\cos\theta,
(\sinh\eta)\sin\theta, 0 \right) .
\end{equation}
The rotation matrix which performs this operation is
\begin{equation} \label{rot44}
R(\theta) = \pmatrix{1 & 0 & 0 & 0 \cr
            0 & \cos\theta & -\sin\theta & 0 \cr
            0 & \sin\theta & \cos\theta & 0 \cr
            0 & 0 & 0 & 1 } .
\end{equation}
Instead of this rotation, we propose to obtain this four-vector by
boosting the four-momentum of Eq.(\ref{4b}).  The boost matrix in this
case is
\widetext
\begin{equation}\label{boost2}
B_{2} = \pmatrix{\cosh\lambda & -\sin(\theta/2)\sinh\lambda  &
              \cos(\theta/2)\sinh\lambda  & 0\cr
-\sin(\theta/2)\sinh\lambda & 1 + \sin^{2}(\theta/2)(\cosh\lambda - 1)
              & -\sin\theta \sinh^{2}(\lambda/2) & 0 \cr
\cos(\theta/2) \sinh\lambda & -\sin\theta \sinh^{2}(\lambda/2) &
              1 + \cos^{2}(\theta/2)(\cosh\lambda - 1) & 0 \cr
0 & 0 & 0 & 1} .
\end{equation}
\narrowtext
\noindent with
\begin{equation}\label{lambda}
\lambda = 2 \tanh^{-1}\left\{[\sin(\theta/2)] \tanh\eta\right\} .
\end{equation}
A detailed calculation of this matrix is given the paper by
Han, {\it et al.}\cite{hkn99}

Next, we boost the four-momentum of Eq.(\ref{4c}) to that of Eq.(\ref{4a}).
The particle is again at rest.  The boost matrix is
\begin{equation}\label{boost3}
  B_{3} = R(\theta) B_{1}^{-1} R(-\theta)
\end{equation}

The net result of these transformations is
\begin{equation}\label{net}
B_{3}~B_{2}~B_{1} .
\end{equation}
This leaves the initial four-momentum of Eq.(\ref{4a}) invariant.  Is it
going to be an identity matrix?  The answer is No.  The result of
the matrix multiplications is
\begin{equation}\label{omega}
R(\Omega) = \pmatrix{1 & 0 & 0 & 0 \cr 0 & \cos\Omega & -\sin\Omega & 0 \cr
   0 & \sin\Omega & \cos\Omega & 0 \cr 0 & 0 & 0 & 1} ,
\end{equation}
with
\begin{equation}\label{omeg}
\Omega = 2\, \sin^{-1}\left\{{(\sin\theta)\sinh^{2}(\eta/2) \over
\sqrt{\cosh^{2}\eta - \sinh^{2}\eta \sin^{2}(\theta/2)} }\right\} .
\end{equation}
This matrix performs a rotation around the $y$ axis and leaves the
four-momentum of Eq.(\ref{4a}) invariant.  This rotation is an
element of Wigner's little group whose transformations leave the
four-momentum invariant.  This is precisely the Wigner rotation.

Indeed, Wigner's little group is the maximum subgroup of the Lorentz
group whose transformations leave the four-momentum of a given particle
invariant.  The Wigner rotation is associated with the little group
for a particle at rest.  Then, how about the little group which leaves
the four-momentum $P_{b}$ of Eq.(\ref{4b})?

As Wigner noted, this four-vector can be brought to $P_{a}$ of
Eq.(\ref{4a}) by the inverse of the matrix of $B_{1}$.  Then the rotation
matrix
\begin{equation} \label{wrot44}
R(\Theta) = \pmatrix{1 & 0 & 0 & 0 \cr
            0 & \cos\Theta & -\sin\Theta & 0 \cr
            0 & \sin\Theta & \cos\Theta & 0 \cr
            0 & 0 & 0 & 1 } ,
\end{equation}
leaves the four-momentum $P_{a}$ invariant.  Thus, the transformation
\begin{equation}\label{lgb1}
B_{1}~R(\Theta)~B_{1}^{-1}
\end{equation}
will leave the four-momentum $P_{b}$ invariant.  Clearly the rotation
of Eq.(\ref{wrot44}) is a Wigner rotation~\cite{hks86jm}.

\begin{figure}[thb]
\begin{center}
\epsfysize=80mm \epsffile{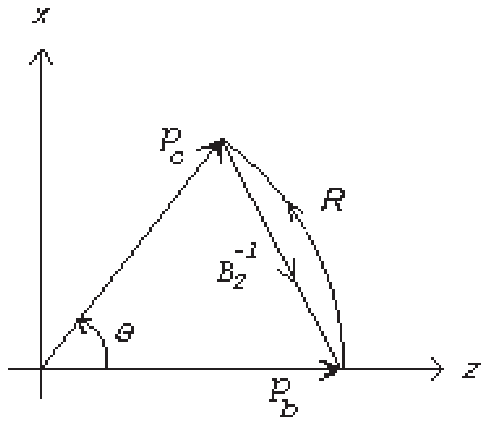} \caption{Lorentz-boosted
rotation.  If the particle moves along the $z$ direction, it can
be brought to its rest frame by the inverse of the boost matrix
$B_{1}$.  When it is at rest, we can rotate the system without
changing while its momentum.  Under this rotation, the spin of
the particle will change its direction.  The particle can then be
brought to its initial state by the boost matrix $B_{1}$. The
initial four-momentum can also be rotated by first as indicated
in this figure.  It can then be boosted back to its initial
momentum state. The net result is a matrix which does not change
the momentum.  This can also be achieved by a Lorentz-boosted
rotation around the $y$ axis.} \label{wr22}
\end{center}
\end{figure}
According to Wigner's definition based on the $O(3)$-like little group,
both  $R(\Omega)$ of Eq.(\ref{omega}) and $R(\Theta)$ are Wigner
rotations.  In the case of $R(\Omega)$, the rotation angle is determined
by the kinematical parameters $\eta$ and $\theta$.  On the other hand,
the angle $\Theta$ is arbitrary.  In order to see those two seemingly
different rotations are equivalent, we shall convert the kinematics
of $\Theta$ into that of $\Omega$.

For this purpose, we note first that there are two ways of transforming
$P_{b}$ to $P_{c}$.  One is the rotation of Eq.(\ref{rot44}), and
the other is the boost $B_{2}$ of Eq.(\ref{boost2}).  Thus the transformation
\begin{equation}\label{lgb2}
B_{2}^{-1}~R(\theta)
\end{equation}
will leave the four-vector $P_{b}$ invariant.  If we put the
restriction that the transformations
of Eq.(\ref{lgb1}) and Eq.(\ref{lgb2}) be equal:
\begin{equation}
B_{1}~R(\Theta)~B_{1}^{-1} = B_{2}^{-1}~R(\theta) ,
\end{equation}
then the result is
\begin{equation}
R(\theta)~R(\Theta) = B_{3}~B_{2}~B_{1} ,
\end{equation}
or
\begin{equation}
\Theta = \Omega - \theta .
\end{equation}

In 1986~\cite{hks86jm} and 1999~\cite{hkn99}, Han {\it et al.} performed
exactly the same calculation
using two-by-two formalism applicable to the Jones matrix formalism
in polarization optics.  They of course used the correspondence
between the $O(2,1)$ and $Sp(2)$ groups.  The rotation of
Eq.(\ref{rot44}) is translated into a two-by-two rotation matrix
with $\theta$ replaced by $\theta/2$.

The two-by-two  squeeze matrix corresponding to the boost
matrix $B_{1}$ of Eq.(\ref{boost1}) is
\begin{equation}
S_{1} = \pmatrix{e^{\eta/2} & 0 \cr 0 & e^{-\eta/2} } .
\end{equation}
The two-by-two rotation matrix corresponding to the four-by-four
rotation matrix of Eq.(\ref{rot44}) is
\begin{equation}\label{rot22}
R(\theta) = \pmatrix{\cos(\theta/2) & -\sin(\theta/2) \cr
          \sin(\theta/2) & \cos(\theta/2) }  .
\end{equation}
After the matrix multiplication, the squeeze matrix $S_{2}$
corresponding to $B_{2}$ of Eq.(\ref{boost2}) becomes~\cite{hkn99}
\widetext
\begin{equation}
S_{2} = \pmatrix{\cosh(\lambda/2) - \sin(\theta/2) \sinh(\lambda/2) &
\cos(\theta/2) \sinh(\lambda/2) \cr
\cos(\theta/2) \sinh(\lambda/2) &
\cosh(\lambda/2) + \sin(\theta/2) \sinh(\lambda/2)} .
\end{equation}
This is a matrix which squeezes along the direction which makes
the angle $(\pi + \theta)/2$ with the $z$ axis.  The two-by-two
squeeze matrix corresponding to $B_{3}$ of Eq.(\ref{boost3}) is
\begin{equation}
S_{3} = \pmatrix{\cosh(\eta/2) - \cos\theta \sinh(\eta/2) &
         - \sin\theta \sinh(\eta/2) \cr
         - \sin\theta \sinh(\eta/2) &
          \cosh(\eta/2) + \cos\theta \sinh(\eta/2) } .
\end{equation}
\narrowtext
Now the matrix multiplication $S_{3} S_{2} S_{1}$ corresponds to the
closure of the kinematical triangle given in Fig.~\ref{wr11}.
The result is
\begin{equation}
S_{3}S_{2}S_{1} = \pmatrix{\cos(\Omega/2) & -\sin(\Omega/2) \cr
                  \sin(\Omega/2) & \cos(\Omega/2) } ,
\end{equation}
where $\Omega$ is given in Eq.(\ref{omeg}).

Even though the above two-by-two formalism is contained in a paper
on polarization optics~\cite{hkn99}, it is applicable to other
subjects of physics having the $Sp(2)$ symmetry.  Cavity optics is
a case in point.  It is an extension of lens optics governed by
the two-by-two matrices of $Sp(2)$.

Before discussing cavities, let us go back to the definition of
the Wigner rotation.  In his original paper, Wigner introduced the
$O(3)$ group as the subgroup of the rotation group which leaves the
four-momentum of a rest particle invariant.  In the kinematical
configuration of Fig.~\ref{wr11} and in Eq.(\ref{net}), the net
transformation leaves the four-momentum $P_{a}$ of Eq.(\ref{4a})
invariant.  It is a rotation matrix.

With this point in mind, we can write Eq.(\ref{lgb1}) as
\begin{equation}\label{sqrot}
 \pmatrix{e^{\eta/2} & 0 \cr 0 & e^{-\eta/2}}
\pmatrix{\cos(\Theta/2) & -\sin(\Theta/2) \cr
   \sin(\Theta/2) & \cos(\Theta/2)}
 \pmatrix{e^{-\eta/2} & 0 \cr 0 & e^{\eta/2}} .
\end{equation}
Now, these three matrices can be combined into one matrix:
\begin{equation}
\pmatrix{\cos(\Theta/2) & -e^{\eta} \sin(\Theta/2) \cr
  e^{-\eta} \sin(\Theta/2) & \cos(\Theta/2)} .
\end{equation}
If we repeat the same operation $N$ times, the angle $\Theta$
becomes $N\Theta$.

We are now ready to discuss what is happening in a laser cavity.
Let us consider for simplicity a cavity consisting of two identical
concave mirrors separated by by a distance $d$.  Then the $ABCD$ matrix
for a round trip of one beam is
\begin{equation}\label{abcd}
  \pmatrix{1 & 0 \cr -2/R & 1} \pmatrix{1 & d \cr 0 & 1}
  \pmatrix{1 & 0 \cr -2/R & 1} \pmatrix{1 & d \cr 0 & 1},
\end{equation}
where $R$ is the radius of the mirror.  This form is quite familiar
to us from the laser literature~\cite{yari75,haus84,hawk95}.  However,
the crucial question is what happens when this process is repeated
many times.  This question was also addressed in the literature.
For this purpose, Haus replaces one of the concave mirrors with a
flat mirror and repeats the process in order to complete the
cycle~\cite{haus84}.
   We note that Haus's procedure is equivalent to starting the cycle
from the midpoint between the mirrors.  This procedure can be
simplified if we introduce a group theoretical notion of equivalent
class.  This procedure is simple.  We translate the system by $d/2$
using a translation matrix.  We thus write the $ABCD$ matrix of
Eq.(\ref{abcd}) as
\widetext
\begin{equation}
\pmatrix{1 & -d/2 \cr 0 & 1}
\left[\pmatrix{1 - d/R &   d - d^{2}/2R  \cr -2/f & 1 - d/R}\right]^{2}
\pmatrix{1 & d/2 \cr 0 & 1}.
\end{equation}
Furthermore,
\begin{equation}
\pmatrix{1 - d/R &   d - d^{2}/2R  \cr -2/R & 1 - d/R}
= \pmatrix{\sqrt{d} & 0 \cr 0 & 1/\sqrt{d}}
\pmatrix{1 - d/R &   1 - d/2R  \cr -2d/R & 1 - d/R}
\pmatrix{1/\sqrt{d} & 0 \cr 0 & \sqrt{d}} .
\end{equation}
The purpose of this decomposition was to write the matrix in the middle
in terms of dimensionless quantities.

Now, the $ABCD$ matrix of Eq.(\ref{abcd}) can be written as
\begin{equation}
\pmatrix{1 & -d/2 \cr 0 & 1}
\pmatrix{\sqrt{d} & 0 \cr 0 & 1/\sqrt{d}}
\left[\pmatrix{1 - d/R &   1 - d/2R  \cr -2d/R & 1 - d/R}\right]^{2}
\pmatrix{1/\sqrt{d} & 0 \cr 0 & \sqrt{d}}
\pmatrix{1 & d/2 \cr 0 & 1}.
\end{equation}
If the beam makes $N$ round trips, the $ABCD$ matrix becomes
\begin{equation}
\pmatrix{1 & -d/2 \cr 0 & 1}
\pmatrix{\sqrt{d} & 0 \cr 0 & 1/\sqrt{d}}
\left[\pmatrix{1 - d/R &  1 - d/2R  \cr -2d/R & 1 - d/R}\right]^{2N}
\pmatrix{1/\sqrt{d} & 0 \cr 0 & \sqrt{d}}
\pmatrix{1 & d/2 \cr 0 & 1}.
\end{equation}
\narrowtext
Thus, we can thus decompose this expression into a core matrix $C$, and
the escort matrix $E$ and its inverse $E^{-1}$ in the following manner.
\begin{equation}
E~C^{2N}~E^{-1} ,
\end{equation}
with
\begin{eqnarray}
&{}&  C = \pmatrix{1 - d/R &  1 - d/2R  \cr
         -2d/R & 1 - d/R},  \nonumber \\[2mm]
&{}&  E = \pmatrix{1 & -d/2 \cr 0 & 1}
\pmatrix{\sqrt{d} & 0 \cr 0 & 1/\sqrt{d}} .
\end{eqnarray}
With this expression, we can concentrate on the core matrix $C$,
and write this in the form
\begin{equation}
C = \pmatrix{\cos\phi & -e^{\xi}\sin\phi  \cr
    e^{-\xi} \sin\phi  & \cos\phi} ,
\end{equation}
with
\begin{equation}
\cos\phi = 1 - {d \over R}, \quad
e^{2\xi} = {R \over 2d} - {1 \over 4} .
\end{equation}
Here both $d$ and $R$ are positive, and the restriction on them is
that $d$ be greater than $2R$.  This is the stability condition
frequently mentioned in the literature~\cite{haus84,hawk95}.

Let us next write the core matrix as
\begin{equation}
 C = \pmatrix{e^{\eta/2}  & 0 \cr  0  & e^{-\eta/2}}
    \pmatrix{\cos\phi & -\sin\phi  \cr \sin\phi  &
    \cos\phi}
    \pmatrix{e^{-\eta/2} & 0 \cr 0 & e^{\eta/2}}  .
\end{equation}
Here, a rotation matrix is sandwiched between a squeeze matrix and its
inverse.  This expression is exactly of the form of Eq.(\ref{sqrot})
for the Wigner rotation.  In the above expression also, the rotation
matrix in the middle is the Wigner rotation matrix.

If the light beam makes one cycle, the effect is $C^{2}$, and the
its expression is
\begin{equation}
 C = \pmatrix{e^{\eta/2}  & 0 \cr  0  & e^{-\eta/2}}
    \pmatrix{\cos(2\phi) & -\sin(2\phi)  \cr \sin(2\phi)  &
    \cos(2\phi)}
    \pmatrix{e^{-\eta/2} & 0 \cr 0 & e^{\eta/2}}  .
\end{equation}
Indeed, the beam makes a Wigner rotation of $2\phi$ when it completes
one cycle.

If the light beam makes $N$ round trips, we have to compute $C^{2N}$,
and the result is
\begin{equation}
C^{2N} = \pmatrix{e^{\eta/2}  & 0 \cr  0  & e^{-\eta/2}}
 \pmatrix{\cos(2N\phi) & -\sin(2N\phi) \cr \sin(2N\phi) &
 \cos(2N\phi)}
 \pmatrix{e^{-\eta/2}  & 0 \cr  0  & e^{\eta/2}}  ,
\end{equation}
or
\begin{equation}
C^{2N} = \pmatrix{\cos(2N\phi) & -e^{\eta} \sin(2N\phi) \cr
    e^{-\eta} \sin(2N\phi) &  \cos(2N\phi)}  .
\end{equation}

In this paper, we noted first that the matrices in lens/mirror optics
can be formulated in terms of the three-parameter $Sp(2)$ group.  We
exploited the isomorphism between $Sp(2)$ and $SO(2,1)$ which is the
Lorentz group for the particles moving in a two-dimensional plane.
The Wigner rotation, with its group theoretical origin, manifests
itself in special relativity and optical sciences including cavity
optics.  It is gratifying to note that laser beams perform many
Wigner rotations before they leave the cavity.

In this paper, we considered only the simplest cavity consisting of
two identical mirrors.  We note that there are more general approaches
for cavities consisting of two different mirrors~\cite{yari75}.  It
would be an interesting project to exploit the Lorentz-group content of
this and other general cases.

\end{document}